\documentclass[prb,aps,twocolumn,amsmath,amssymb,floatfix,superscriptaddress]{revtex4}
\usepackage[dvips]{graphics}
\usepackage{color}
\definecolor{dred}{rgb}{0,0,0.6}
\usepackage{amsmath}
\usepackage{graphicx}
\usepackage{xcolor}
\usepackage{amssymb}

\begin{document}

\title{Longer disordered rings conduct better}

\author{Moumita Patra}

\email{moumita.patra19@gmail.com}

\affiliation{School of Physical Sciences, Indian Association for the Cultivation
of Science, 2A \& 2B Raja S. C. Mullick Road, Kolkata 700032, India}

\begin{abstract}
Weak disorder can make a longer quantum ring conduct better.
Using a tight-binding Green's function approach, we show that
disorder lifts interference-induced transmission zeros,
producing contact-controlled anomalous conductance scaling.
At zero energy, an initial enhancement gives way to apparent
saturation or decay, depending on contact geometry.
At selected nonzero degenerate energies, weak disorder supports
sustained growth in the upper envelope of the even-size sequence
for asymmetric contacts, whereas symmetric contacts yield
decreasing conductance. An analytical treatment captures the
growth and its curvature without fitted parameters.
In this weak-disorder regime, the injected response remains
spatially extensive, with neither exponential conductance
suppression nor participation-ratio saturation observed over
the investigated size range. These findings identify
disorder-modified interference as a mechanism for conductance
growth with system size despite random disorder.
\end{abstract}

\maketitle

Anderson localization establishes that interference between waves
scattered by random disorder can suppress diffusion without interactions
or inelastic processes~\cite{r01,r02,r03}. In one-dimensional systems
with uncorrelated on-site disorder and short-range hopping, the typical
conductance decreases exponentially with system
length~\cite{r04,r05,r06,r07}. Spectral anomalies and conductance
fluctuations refine this description and make the distinction between
arithmetic and logarithmic averages essential for finite-size
scaling~\cite{r08,r09,r10,r11,r12,r13}. Observations in photonic
lattices and ultracold atomic systems further establish localization
as a general wave phenomenon~\cite{r14,r15,r16,r17}. Nevertheless,
a finite coherent conductor can already exhibit vanishing transmission
in the absence of disorder. Whether random scattering can reverse
the length dependence of conductance by weakening such an
interference blockade is therefore a distinct transport question.

In mesoscopic devices, transmission between reservoirs makes
contacts and boundary conditions integral to
conductance~\cite{r18,r19,r20,r21,r22,r23}, allowing phase-sensitive
transport beyond what the isolated spectrum alone
predicts~\cite{r24,n01,n02}. Quantum rings provide a minimal
two-path interferometer~\cite{r26,r28,r29}, with phase-sensitive
transport demonstrated in metallic rings, semiconductor
interferometers, and nanowire networks~\cite{r30,r32,r33,n19}.
Destructive interference generates antiresonances closely
related to Fano transmission zeros~\cite{r34,r35,r37,r38},
whose occurrence depends on contact geometry and electrode
hybridization~\cite{r41}. Different couplings of eigenstate
combinations to a continuum can produce broad or narrow
resonances and decoupled states~\cite{r44,r46,r47}. Thus,
isolated-ring degeneracy alone does not determine open-system
transmission.

\begin{figure}
{\centering \resizebox*{8cm}{3.25cm}{\includegraphics{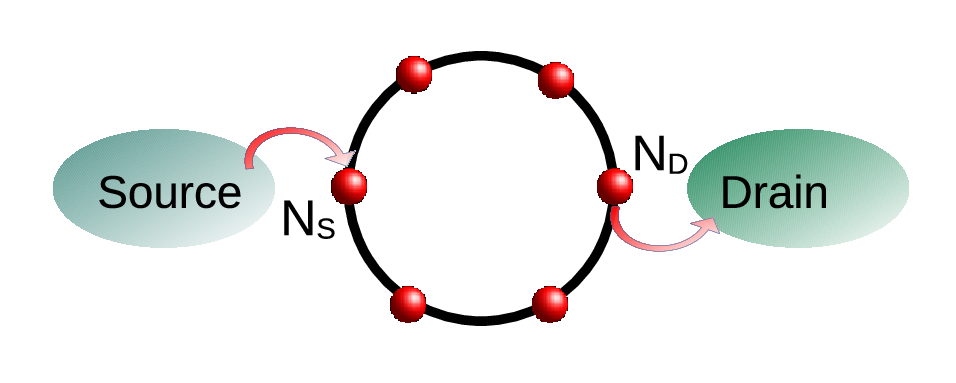}}\par}
\caption{(Color online) Schematic of a tight-binding quantum ring coupled
to source and drain electrodes at sites $N_S$ and $N_D$, respectively.
Random on-site disorder is restricted to the ring.}
\label{f1}
\end{figure}

Recent advances have enabled experimental control of quantum
interference in nanoscale transport~\cite{n03}, including direct
observation and electrochemical manipulation of antiresonances in
molecular junctions~\cite{n04,n05}. In flat-band systems, the
connection between compact localized states and Fano resonances
underlies unconventional localization behavior~\cite{n08,n09,n10,n11,n12}.
Interference-induced confinement has been observed in photonic
Aharonov--Bohm cages~\cite{n13}, while correlated disorder can lift
such confinement, producing inverse Anderson transitions in
theoretical models, ultracold atoms, and electrical
circuits~\cite{n14,n15,r54,n16}. Transport enhancement has also
been explored in disordered spin and coherent networks~\cite{n18,n20},
systems with long-range hopping or cavity-mediated
coupling~\cite{r55,r56,r57}, and twisted bilayer graphene~\cite{n17}.
These developments raise a complementary question: can a simple
nearest-neighbor ring with uncorrelated on-site disorder exhibit
increasing disorder-averaged conductance with size at fixed
disorder strength?

In this Letter, we demonstrate contact-controlled anomalous
conductance scaling in disordered quantum rings coupled
to two electrodes. Static disorder modifies ring
antiresonances~\cite{r48} while preserving phase coherence,
unlike inelastic scattering~\cite{r58,r59}. Within a
tight-binding nonequilibrium Green's function (NEGF)
formalism, we show that asymmetric contacts support an
initial conductance increase at zero energy, followed
by apparent saturation or decay depending on the contact
geometry. At selected nonzero degenerate eigenenergies
of the ordered ring, weak on-site disorder produces
sustained conductance growth in the upper envelope of
the full even-size sequence. Energy-resolved transmission
and an analytical treatment identify the mechanism:
disorder lifts the destructive-interference blockade
imposed by asymmetric contacts. Symmetric contacts
instead yield decreasing conductance. At the selected
nonzero energies and weak disorder strengths, both trends
coexist with a spatially extensive injected response;
neither exponential conductance suppression nor
participation-ratio saturation is observed over the
investigated size range. These results identify
disorder-modified interference as a mechanism that allows
longer disordered rings to transmit more efficiently,
without correlated disorder, long-range hopping, or
dephasing.

To investigate this behavior, we consider an $N$-site ring described
by the noninteracting tight-binding Hamiltonian
\begin{equation}
H_{\mathrm C}=\sum_{i=1}^{N}\epsilon_i c_i^\dagger c_i
+v\sum_{i=1}^{N}(c_i^\dagger c_{i+1}+\mathrm{H.c.}),
\end{equation}
with periodic boundary conditions $c_{N+1}=c_1$. The on-site energies
are independently drawn from $[-\Delta W/2,\Delta W/2]$.
Two semi-infinite one-dimensional electrodes couple to ring sites
$N_S$ and $N_D$ through hopping amplitudes $\tau_S$ and $\tau_D$.
Within the NEGF formalism, the retarded Green function is
$\mathcal G^r=[(E+i0^+)I-H_{\mathrm C}-\Sigma_S^r-\Sigma_D^r]^{-1}$,
where
$\Sigma_\alpha^r=|\tau_\alpha|^2g_0^r
|N_\alpha\rangle\langle N_\alpha|$.
For electrode hopping $t_0$ and on-site energy $\epsilon_0$, the
surface Green function within the electrode band is
$g_0^r=[E-\epsilon_0-i\sqrt{4t_0^2-(E-\epsilon_0)^2}]/(2t_0^2)$.
The zero-temperature linear conductance follows as~\cite{r20,r23}
\begin{equation}
G(E_F)=G_0\operatorname{Tr}
[\Gamma_S\mathcal G^r\Gamma_D\mathcal G^a]_{E=E_F},
\end{equation}
where $\Gamma_\alpha=i(\Sigma_\alpha^r-\Sigma_\alpha^a)$,
$\mathcal G^a=(\mathcal G^r)^\dagger$, and $G_0$ is the conductance
quantum. Disorder averages are evaluated over 1000 independent
configurations.

For the calculations below, we set the electrode and ring
hopping amplitudes to $t_0=4\,\mathrm{eV}$ and
$v=1\,\mathrm{eV}$, respectively, unless stated otherwise.
We use equal ring--electrode couplings,
$\tau_c=\tau_S=\tau_D$, and define the conductance quantum
as $G_0=2e^2/h$. All on-site energies vanish in the ordered
system, and the source remains attached to site $N_S=1$
throughout.

\begin{figure}
{\centering \resizebox*{8.4cm}{3.2cm}{\includegraphics{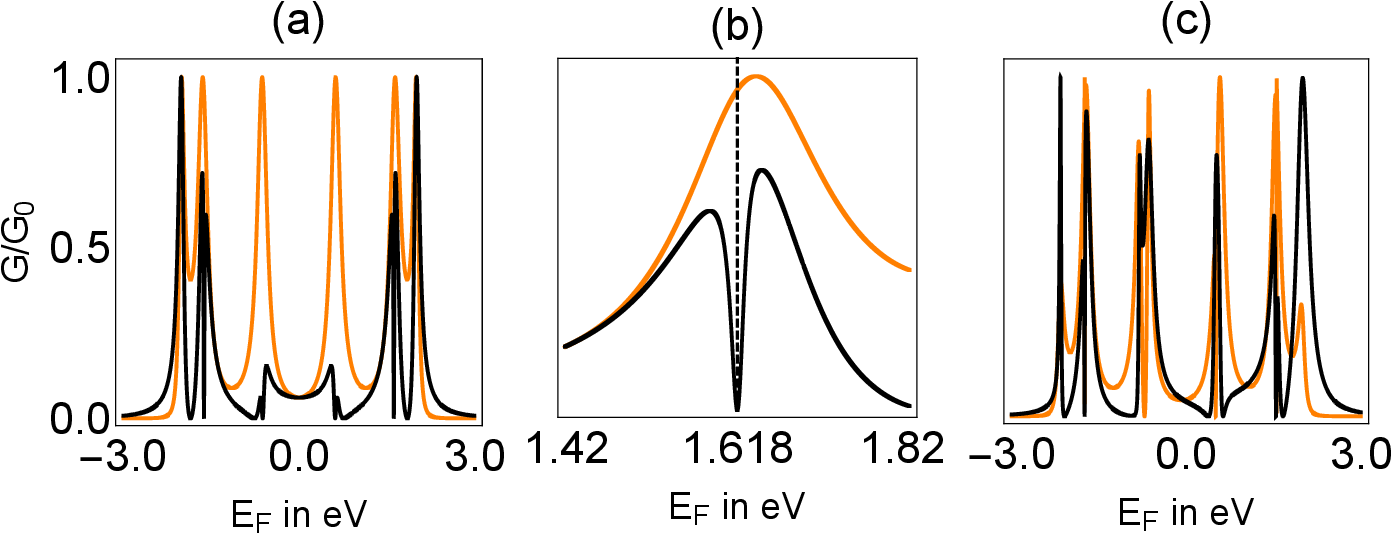}}
\resizebox*{2.8cm}{2.8cm}{\includegraphics{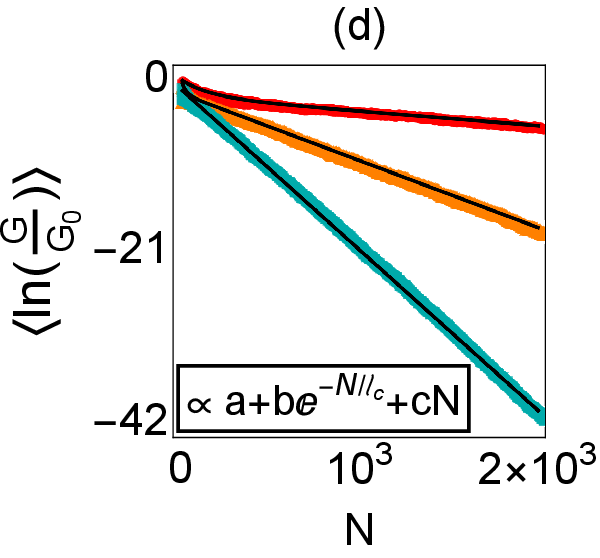}}
\resizebox*{2.8cm}{2.8cm}{\includegraphics{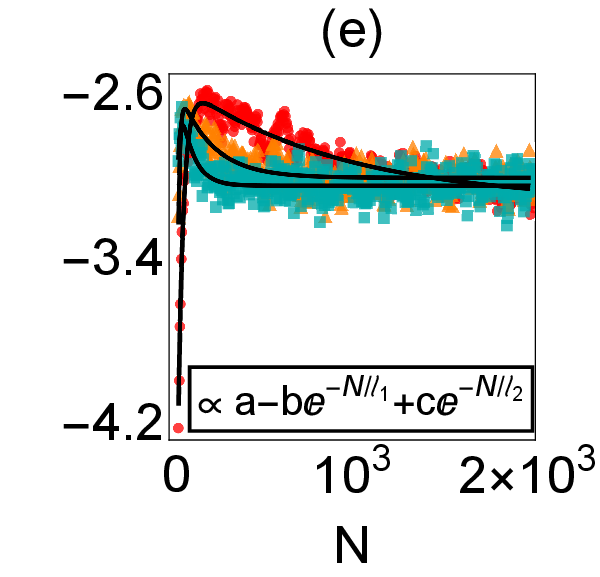}}
\resizebox*{2.8cm}{2.8cm}{\includegraphics{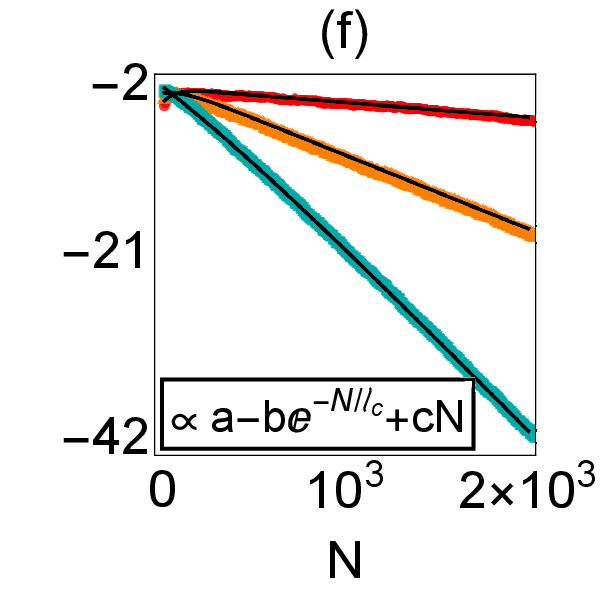}}\par}
\caption{(Color online) (a) Conductance spectra of an ordered
ten-site ring with symmetric (orange) and maximally asymmetric
(black) contacts. (b) Enlargement near
$E_F=1.618034\,\mathrm{eV}$ (vertical dashed line).
(c) Corresponding spectra for one disorder realization with
$\Delta W=1\,\mathrm{eV}$.
(d)--(f) $\langle\ln(G/G_0)\rangle$ versus $N$ at $E_F=0$
for symmetric, maximally asymmetric, and minimally asymmetric
($N_D=N/2+2$) contacts, respectively.
Red circles, orange triangles, and teal squares denote
$\Delta W=0.5$, $1.0$, and $1.5\,\mathrm{eV}$, respectively.
Black curves denote fits, with the corresponding scaling forms
indicated in each panel and discussed in the text.
The ring--electrode couplings are
$\tau_S=\tau_D=1\,\mathrm{eV}$ throughout.}
\label{f2}
\end{figure}

Interference between the arms of a quantum ring can reverse
the usual conductance decrease with size, depending on Fermi
energy, contact geometry, and ring--electrode coupling.
Figure~\ref{f2}(a) compares an ordered ten-site ring with symmetric
($N_D=6$) and maximally asymmetric ($N_D=10$) contacts.
The isolated-ring levels are twofold degenerate except at the
band edges. For a pair with wave vectors $\pm k_m$,
$k_m=2\pi m/N$, the contact separation $d=N_D-1$ determines
the interference condition: when $\sin(k_md)=0$, one linear
combination decouples from both electrodes; otherwise,
destructive interference produces a transmission zero at
the isolated-ring eigenenergy. Symmetric contacts yield six
distinct resonances, whereas maximal asymmetry introduces
antiresonances. Near
$E_m=2v\cos(\pi/5)\simeq1.618034\,\mathrm{eV}$, the zero
remains pinned to $E_m$, while coupling shifts the neighboring
resonance maxima [Fig.~\ref{f2}(b)]. Disorder modifies this
interference, shifting or lifting the zeros [Fig.~\ref{f2}(c)].

At $E_F=0$, a degenerate eigenenergy for $N=4m$, contact
geometry controls anomalous size scaling. We analyze
$y(N)=\langle\ln(G/G_0)\rangle$, with typical conductance
$G_{\mathrm{typ}}=G_0e^{y(N)}$. All fits use
$N=12,16,\ldots,2040$; parameter sets correspond to
$\Delta W=0.5$, $1.0$, and $1.5\,\mathrm{eV}$, respectively.
For symmetric contacts, $y(N)$ decreases throughout
[Fig.~\ref{f2}(d)]. The empirical form
$y(N)\simeq a+b e^{-N/\ell_c}+cN$ gives
$(a,b,c,\ell_c)=
(-2.44589,2.12686,-0.00178153,168.018)$,
$(-2.05529,1.96726,-0.00822861,27.366)$, and
$(-1.49566,1.99807\times10^{-5},-0.0197333,30.826)$.
For the strongest disorder, the exponential correction is
negligible and its scale poorly constrained. Beyond the
transient, approximately linear decay of $y(N)$ corresponds
to exponential suppression of $G_{\mathrm{typ}}$.

Maximal asymmetry instead yields an initial conductance
increase, a maximum, and a weakly size-dependent regime
[Fig.~\ref{f2}(e)], described by
\begin{equation}
y(N)\simeq a-b e^{-N/\ell_1}+c e^{-N/\ell_2},
\qquad b,c>0,\quad 0<\ell_1<\ell_2.
\end{equation}
The fitted parameters $(a, b, c, \ell_1,\ell_2)$ are
$(-3.13416$, $2.3084$, $0.586843$, $30.4141$, $905.814)$,
$(-3.0164$, $2.01625$, $0.459352$, $10.546$, $188.933)$, and
$(-3.05486$, $0.877148$, $0.582792$, $13.2647$, $71.6935)$.
For $N_D=N$, increasing size modifies interference while
preserving a one-bond source--drain path, whose transmission
can survive suppression of the long-arm contribution.
Here $a$ is a fit offset, not an established physical
asymptote.

Minimal asymmetry, $N_D=N/2+2$, also supports an initial
enhancement at lower disorder, followed by sustained decay
[Fig.~\ref{f2}(f)]. Both paths now grow as $N/2\pm1$.
The fit
\begin{equation}
y(N)\simeq a-b e^{-N/\ell_c}+cN,
\qquad b>0,\quad c<0,\quad \ell_c>0,
\end{equation}
yields $(a, b, c,\ell_c)=$
$(-2.65552$, $1.77425$, $-0.00166939$, $64.583)$,
$(-1.96922$, $1.31104$, $-0.00830371$, $110.284)$, and
$(0.763281$, $3.26516$, $-0.0209971$, $662.712)$.
The first two fits retain an initial rise; the third
decreases monotonically. Stronger disorder thus suppresses
the enhancement and steepens the subsequent decay.
These fits quantify finite-size crossovers, not asymptotic
scaling laws.

Increasing the size of a disordered ring can therefore
enhance its typical conductance over a finite interval,
with contact geometry controlling the subsequent crossover.
We next examine nonzero degenerate eigenenergies, where
the enhancement can persist over a broader size range.
Unlike the zero-energy transmission zero within a broader
suppressed region, the selected nonzero-energy antiresonances
lie between resolved peaks. We reduce the ring--electrode
coupling to limit broadening and track how weak disorder
lifts the interference blockade.

\begin{figure}
{\centering \resizebox*{3.8cm}{3.8cm}{\includegraphics{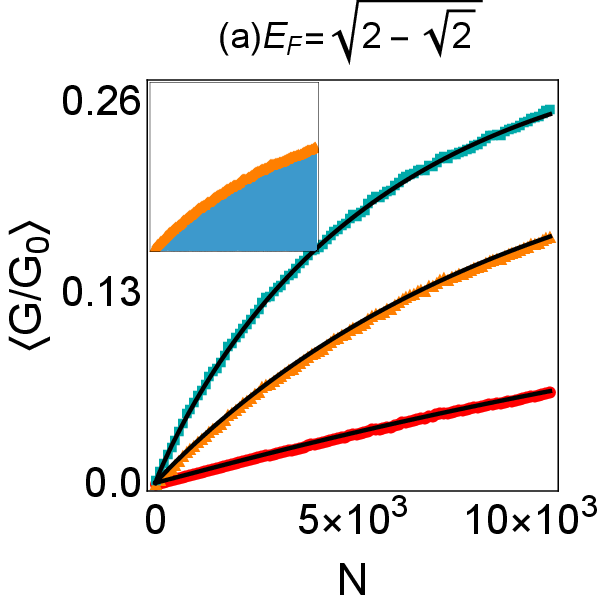}}
\resizebox*{3.8cm}{3.8cm}{\includegraphics{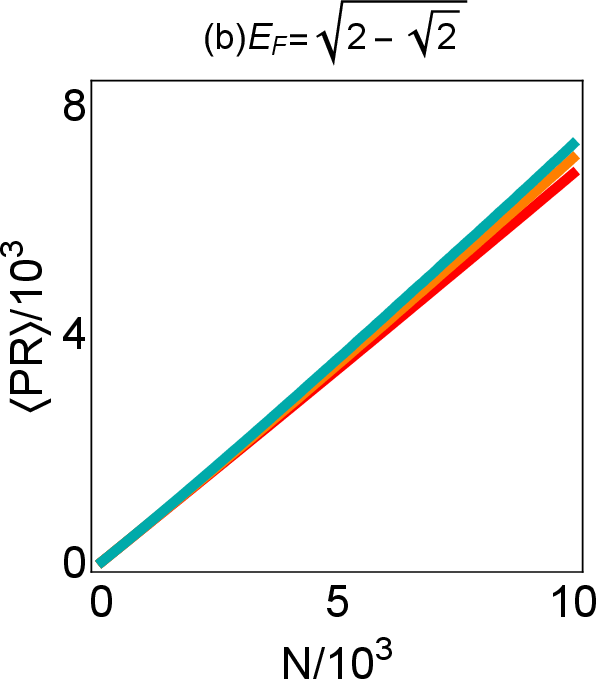}}
\resizebox*{3.8cm}{3.8cm}{\includegraphics{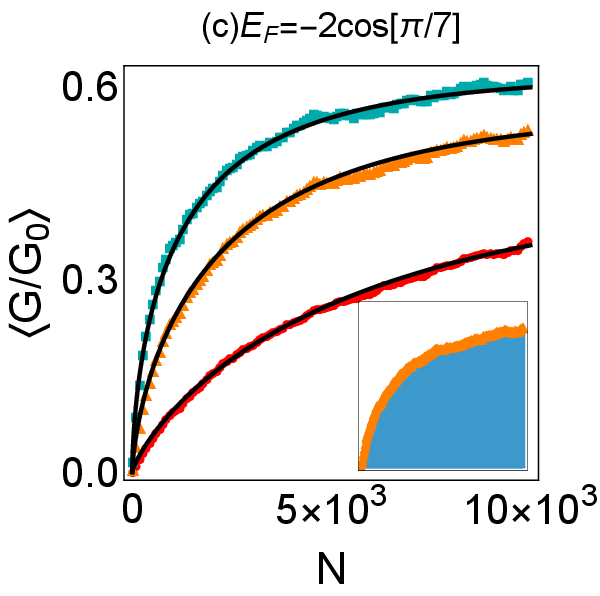}}
\resizebox*{3.8cm}{3.8cm}{\includegraphics{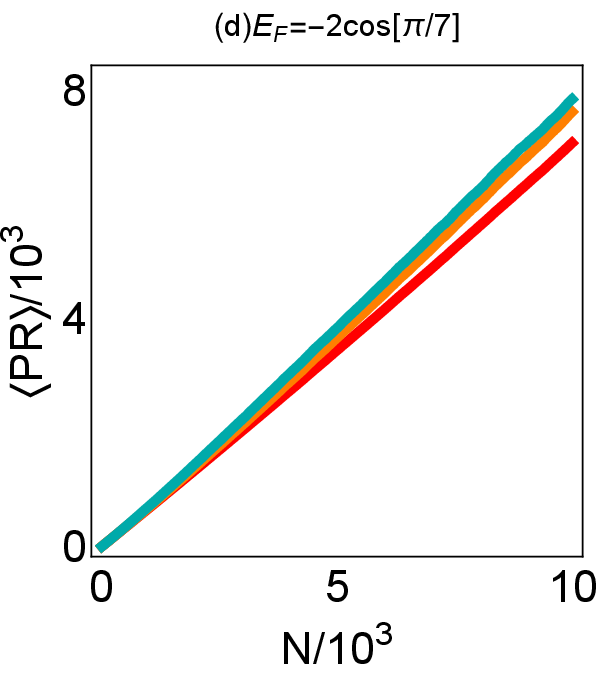}}
\resizebox*{3.8cm}{3.8cm}{\includegraphics{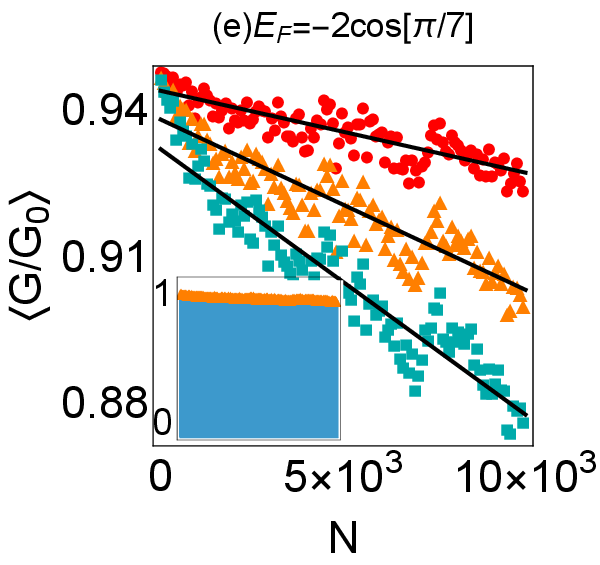}}
\resizebox*{3.8cm}{3.8cm}{\includegraphics{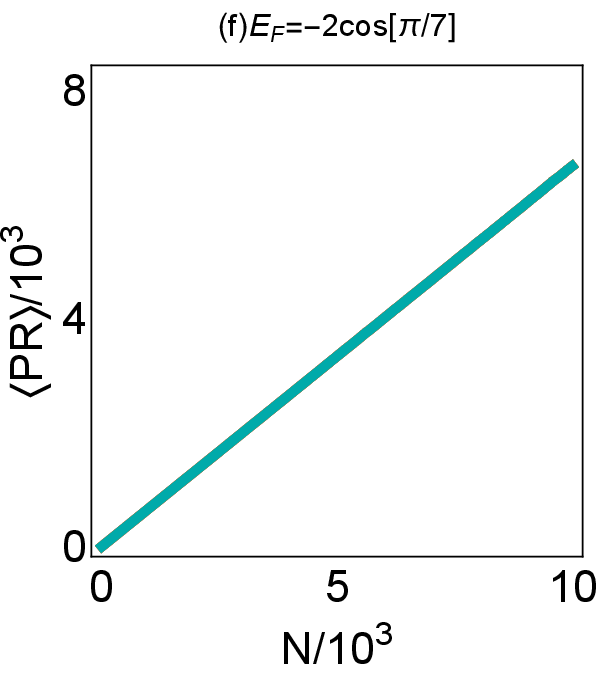}}\par}
\caption{(Color online) Disorder-averaged conductance,
$\langle G/G_0\rangle$, and participation ratio,
$\langle\mathrm{PR}\rangle$, versus ring size $N$.
(a), (b) Asymmetric contacts, $N_D=N/2+2$, at
$E_F=\sqrt{2-\sqrt{2}}\,\mathrm{eV}$;
(c), (d) the same at $E_F=-2\cos(\pi/7)\,\mathrm{eV}$;
(e), (f) symmetric contacts at the latter energy.
The main panels use $N=16m$ in (a), (b) and $N=14m$
in (c)--(f), where $m$ is a positive integer, so that
the selected energies are degenerate eigenenergies of
the isolated ordered rings. Red circles, orange triangles,
and cyan squares denote $\Delta W=0.1$, $0.2$, and
$0.3\,\mathrm{meV}$, respectively. Black curves are
fits described in the text. Insets in (a), (c), (e):
conductance for all even $N$ at
$\Delta W=0.2\,\mathrm{meV}$, with the commensurate
sequences highlighted in orange. Their upper envelopes
retain the trends of the main panels. Inset axis labels
and tick scales match those of the corresponding main
panels, except where indicated in (e).
The ring--electrode coupling is
$0.2\,\mathrm{eV}$ in (a), (b) and
$0.25\,\mathrm{eV}$ in (c)--(f).}
\label{f4}
\end{figure}

We now examine the arithmetic mean $\langle G/G_0\rangle$ to
quantify disorder-averaged conductance growth, whereas
Fig.~\ref{f2} characterizes typical conductance through
$\langle\ln(G/G_0)\rangle$. Growth of the arithmetic
mean does not by itself establish the size dependence
of the typical conductance.

Figure~\ref{f4} demonstrates conductance growth at
$E_F=\sqrt{2-\sqrt{2}}\,\mathrm{eV}$ and
$E_F=-2\cos(\pi/7)\,\mathrm{eV}$, which are common
degenerate eigenenergies of ordered isolated rings with
$N=16m$ and $N=14m$, respectively, where $m$ is a positive
integer. For asymmetric contacts, $N_D=N/2+2$,
$\langle G/G_0\rangle$ increases with ring size at both
energies [Figs.~\ref{f4}(a),(c)], establishing that the
growth is not restricted to the zero-energy degeneracy.
The main panels display these commensurate sequences
for clarity. When all even sizes are included, the insets
for $\Delta W=0.2\,\mathrm{meV}$ show strong conductance
oscillations whose upper envelope follows the commensurate
sequence, highlighted in orange. The anomalous growth
therefore remains evident in the upper envelope of the
full even-size sequence.

At both energies, increasing disorder over
$\Delta W=0.1$--$0.3\,\mathrm{meV}$ further enhances the
conductance. The solid black curves fit the growth
through
\begin{equation}
\langle G/G_0\rangle=a[1-e^{-(N/\tau)^p}],
\end{equation}
where $a$, $\tau$, and $p$ specify the fitted saturation
value, characteristic size, and growth profile. The analytical
treatment in the End Matter derives the leading conductance
growth and captures its curvature without fitted
parameters [Fig. \ref{f6}]. The approach
toward saturation is more pronounced in
Fig.~\ref{f4}(c). At these weak disorder strengths, the
localization length can greatly exceed the ring sizes
considered. The central observation is the sustained increase
of conductance with $N$, driven by disorder-induced weakening
of the interference blockade. This growth persists throughout
the investigated size range, over which the exponential
conductance suppression characteristic of Anderson localization
is not observed.

The microscopic origin of this anomalous growth can be established
analytically. As derived in the End Matter, independent disorder
contributions to the transmission amplitude add in variance,
yielding $\langle G/G_0\rangle\propto N(\Delta W)^2$ at nonzero
degenerate eigenenergies in the weak-disorder limit. At
the band center, the first-order transmission amplitude
vanishes owing to the sublattice structure, resulting instead
in a leading quadratic size dependence,
$\langle G/G_0\rangle\propto N^2(\Delta W)^4$. Repeated scattering
within the nonzero-energy degenerate pair accounts for the
departure from initial linear growth without adjustable parameters.

To complement the conductance scaling, we examine the spatial
extent of the source-injected response.
Figures~\ref{f4}(b),(d) show the ring-restricted
participation ratio,
\begin{equation}
\mathrm{PR}
=\frac{\left[\sum_{i=1}^{N}
|\mathcal G^r_{iN_S}(E_F)|^2\right]^2}
{\sum_{i=1}^{N}|\mathcal G^r_{iN_S}(E_F)|^4}.
\end{equation}
For a single-site source contact, the scattering-state
amplitude within the ring is proportional to
$\mathcal G^r_{iN_S}$, and the common injection factor
cancels in this ratio. The sums are restricted to the
finite ring, while the infinite electrodes remain
accounted for through their self-energies. Thus,
$\mathrm{PR}$ measures the effective number of ring
sites occupied by the source-injected response.
It is calculated for each disorder realization
before averaging.

At both energies, $\langle\mathrm{PR}\rangle$ grows
approximately linearly with $N$ and remains a substantial
fraction of the ring size. The response therefore
occupies an extensive number of sites, with no
participation-ratio saturation indicative of a finite
localization length within the sampled range.
The growing conductance envelope and approximately linear
increase of $\langle\mathrm{PR}\rangle$ together demonstrate
anomalous conductance growth accompanied by a spatially
extensive injected response. Throughout the investigated
size range, neither exponential conductance suppression
nor saturation of the participation ratio is observed.

Additional numerical checks (not shown) indicate that
the disorder-averaged normalized amplitude profile is
approximately uniform across the ring for the even sizes
examined near the selected degenerate energies and
varies little with the drain positions considered.

For symmetric contacts [Fig.~\ref{f4}(e)], the conductance
at $E_F=-2\cos(\pi/7)\,\mathrm{eV}$ instead decreases
approximately linearly,
\begin{equation}
\langle G/G_0\rangle=c+sN,
\end{equation}
with $c\simeq0.94$ and $s<0$ of order $10^{-6}$,
whose magnitude increases with disorder. The decreasing
envelope persists across all even sizes (inset).
Meanwhile, $\langle\mathrm{PR}\rangle$ remains approximately
proportional to $N$, with nearly coincident curves for
the three disorder strengths [Fig.~\ref{f4}(f)].
Thus, the mean conductance decreases with increasing ring size
even though the injected-response participation ratio remains
approximately proportional to $N$.

These results reveal contact-controlled size scaling:
conductance grows for asymmetric contacts and decreases
weakly for symmetric contacts, while the injected response
remains spatially extensive in both configurations.
The central finding is sustained conductance growth
despite random disorder, with no signatures of
Anderson-type localization in the conductance and
participation-ratio scaling over the investigated range.

\begin{figure}
{\centering \resizebox*{3.8cm}{3.8cm}{\includegraphics{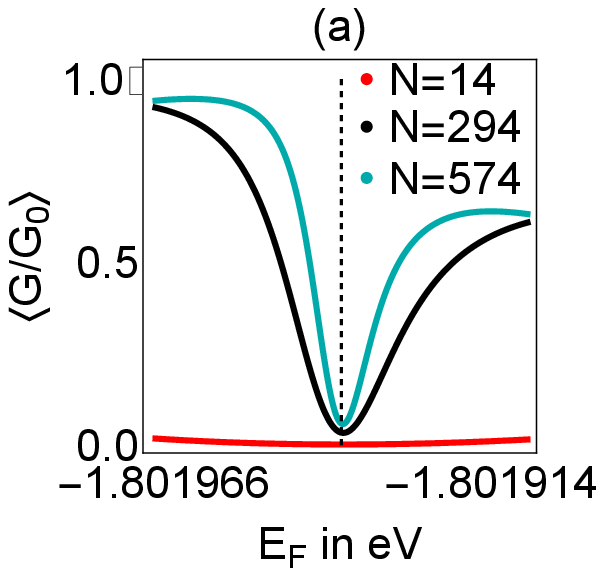}}
\resizebox*{3.8cm}{3.8cm}{\includegraphics{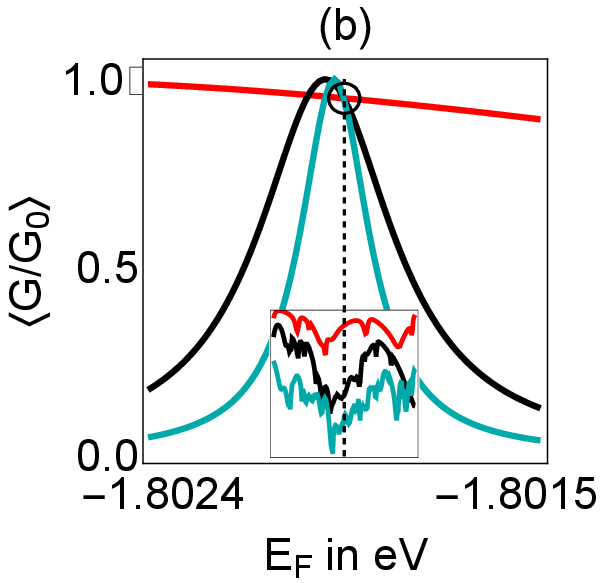}}\par}
\caption{(Color online) $\langle G/G_0\rangle$ near
$E_F^\star=-2\cos(\pi/7)\,\mathrm{eV}$ (dashed line)
for (a) asymmetric contacts, $N_D=N/2+2$, and
(b) symmetric contacts. The inset in (b) enlarges
the circled region over
$|E_F-E_F^\star|\leq0.1\,\mu\mathrm{eV}$;
the plotted conductance values range approximately
from $0.9405$ to $0.9488$.
Parameters: $\Delta W=0.1\,\mathrm{meV}$ and
ring--electrode coupling $\tau_c=0.25\,\mathrm{eV}$.}
\label{f5}
\end{figure}

To understand the anomalous size dependence in
Fig.~\ref{f4}, we examine the conductance spectra near
$E_F^\star=-2\cos(\pi/7)\,\mathrm{eV}$ in
Fig.~\ref{f5}. For asymmetric contacts without disorder,
the conductance vanishes exactly at this energy for
$N=14m$, where $m$ is a positive integer. The remaining
even sizes yield finite conductance with an
$N$-independent upper envelope. Weak random disorder
lifts these transmission zeros, producing the
size-dependent conductance studied here.

For asymmetric contacts [Fig.~\ref{f5}(a)], the selected
energy lies near a conductance minimum. As $N$ increases,
the dip narrows and becomes shallower, increasing the
conductance at the dashed line. For symmetric contacts
[Fig.~\ref{f5}(b)], the selected energy lies near a
conductance peak. Although the peak sharpens and its
maximum remains close to unity, the conductance at the
dashed line decreases with $N$ (inset). These contrasting
spectral changes account for the conductance growth
with asymmetric contacts and its reduction with
symmetric contacts at the selected Fermi energy.

Mesoscopic transport experiments access small energy
scales~\cite{meV1,meV2,meV3}, motivating tests in
phase-coherent quantum rings. Observing the predicted
growth would require thermal broadening, $k_{\mathrm B}T$,
and dephasing-induced broadening, $\Gamma_\phi$, to remain
small compared with the relevant antiresonance width and
its separation from neighboring resonances throughout
the size range studied. Under these conditions,
disorder-induced filling of the transmission minimum
may preserve the anomalous size dependence. The anomalous
growth is therefore expected to persist at sufficiently
low finite temperatures and under weak dephasing,
provided that the relevant interference features remain
resolved.

In conclusion, we demonstrate contact-controlled anomalous
conductance scaling in disordered quantum rings.
At zero energy, asymmetric contacts produce an initial
conductance enhancement followed by apparent saturation
or decay, depending on whether the shortest transport
path remains fixed or grows with ring size.
At selected nonzero degenerate energies, weak disorder
lifts destructive-interference transmission zeros,
yielding sustained growth in the upper envelope of the
even-size sequence. An analytical treatment captures
this growth and its curvature without fitted parameters.
In this regime, asymmetric and symmetric contacts yield
increasing and weakly decreasing conductance, respectively,
while the injected response remains spatially extensive.
Neither exponential conductance suppression nor
participation-ratio saturation is observed over the
investigated size range. These results identify
disorder-modified interference as a mechanism for
conductance growth with system size and show how
contact geometry and system parameters control its
persistence and crossover to decay.

The author expresses her sincere gratitude to
Prof. Santanu K. Maiti for his invaluable guidance and scientific mentorship.

\section*{End Matter}

\subsection*{Analytical origin of conductance growth}

Weak disorder lifts the interference blockade at a clean-ring
antiresonance. We derive the leading contribution and then
retain repeated scattering within the resonant degenerate
pair to describe the curvature at larger sizes.

\paragraph{Leading weak-disorder contribution.}
Label the source and drain sites by $0$ and $d=N_D-1$.
The isolated-ring energies are $E_m=2v\cos k_m$, with
$k_m=2\pi m/N$. For $\sin(k_md)\neq0$, the clean
transmission vanishes at $E_m$, and
\begin{align}
\mathcal G^{(0)}_{j0}
&=-\frac{\sin[k_m(d-j)]}{\Sigma_S\sin(k_md)},\\
\mathcal G^{(0)}_{dj}
&=-\frac{\sin(k_mj)}{\Sigma_D\sin(k_md)},
\end{align}
where $\Sigma_\alpha$ denotes the retarded contact
self-energy and $\Gamma_\alpha=-2\operatorname{Im}\Sigma_\alpha$.
Independent uniform disorder of width $\Delta W$ satisfies
$\langle\epsilon_j\rangle=0$ and
$\langle\epsilon_j\epsilon_l\rangle=\sigma^2\delta_{jl}$,
with $\sigma^2=(\Delta W)^2/12$.

The first-order transmission amplitude,
\begin{equation}
\mathcal G^{(1)}_{d0}
=\sum_j\mathcal G^{(0)}_{dj}\epsilon_j
\mathcal G^{(0)}_{j0},
\end{equation}
has zero mean but finite variance. Consequently,
\begin{equation}
\begin{split}
\left\langle\frac{G(E_m)}{G_0}\right\rangle
={}&\frac{\sigma^2\Gamma_S\Gamma_D}
{|\Sigma_S\Sigma_D|^2\sin^4(k_md)}
\\
&\times\sum_{j=0}^{N-1}
\sin^2(k_mj)\sin^2[k_m(d-j)]\\
&+O((\Delta W)^4).
\end{split}
\label{eq:EM_variance}
\end{equation}
Away from the band center and edges, the sum is
$N[2+\cos(2k_md)]/8$. For identical electrodes with zero
on-site energy, define $\gamma=\tau_c^2/t_0$. Within
the electrode band,
\begin{equation}
\Sigma(E)=\gamma\left[
\frac{E}{2t_0}-i\sqrt{1-\frac{E^2}{4t_0^2}}
\right],
\end{equation}
so $|\Sigma|=\gamma$ and
$\Gamma=2\gamma\sqrt{1-E^2/(4t_0^2)}$. Equation~
\eqref{eq:EM_variance} becomes
\begin{equation}
\begin{split}
\left\langle\frac{G(E_m)}{G_0}\right\rangle
={}&\frac{N(\Delta W)^2}{24\gamma^2}
\left(1-\frac{E_m^2}{4t_0^2}\right)
\\
&\times\frac{2+\cos(2k_md)}{\sin^4(k_md)}
+O((\Delta W)^4).
\end{split}
\label{eq:EM_growth}
\end{equation}
Independent disorder contributions therefore add in
variance, giving linear growth along sequences that
preserve the eigenenergy and contact phase.

For $N_D=N/2+2$, $d=N/2+1$ and $kN/2=m\pi$, yielding
\begin{equation}
\begin{split}
\left\langle\frac{G(E)}{G_0}\right\rangle
={}&\frac{N(\Delta W)^2}{24\gamma^2}
\left(1-\frac{E^2}{4t_0^2}\right)
\\
&\times\frac{1+E^2/(2v^2)}
{[1-E^2/(4v^2)]^2}
+O((\Delta W)^4).
\end{split}
\label{eq:EM_energy}
\end{equation}
Here $E$ is a nonzero degenerate eigenenergy inside
both bands. This result explains the initial dependence
on size, disorder, energy, and coupling. The expansion
is taken at fixed $N$ and is not uniformly accurate
as $N$ increases.

\paragraph{Repeated scattering within the degenerate pair.}
To capture the curvature, we project onto
$\phi_c(j)=\sqrt{2/N}\cos(kj)$ and
$\phi_s(j)=\sqrt{2/N}\sin(kj)$, retaining the full
projected Green matrix. The disorder projection is
\begin{equation}
V_P=
\begin{pmatrix}
A+B&C\\ C&A-B
\end{pmatrix},
\end{equation}
where
\begin{equation}
(A,B,C)=\frac1N\sum_j\epsilon_j
\bigl(1,\cos 2kj,\sin 2kj\bigr).
\end{equation}
At the nonzero energies considered, these variables
have zero means, vanishing pairwise covariances, and
variances $\sigma^2/N$, $\sigma^2/(2N)$, and
$\sigma^2/(2N)$. Their large-$N$ Gaussian approximation is
\begin{equation}
V_P\simeq\frac{\sigma}{\sqrt N}X,\qquad
X=
\begin{pmatrix}
z_0+z_1/\sqrt2&z_2/\sqrt2\\
z_2/\sqrt2&z_0-z_1/\sqrt2
\end{pmatrix},
\end{equation}
with independent standard normal $z_i$.

Define $\boldsymbol u_S=(1,0)^T$,
$\boldsymbol u_D=(\cos\theta,\sin\theta)^T$,
$\theta=kd$, and
\begin{equation}
M=-\sigma\sqrt N\,X
-2\Sigma(E_m)
\left(\boldsymbol u_S\boldsymbol u_S^T+
\boldsymbol u_D\boldsymbol u_D^T\right).
\end{equation}
At $E=E_m$, the projected Green matrix is $NM^{-1}$,
giving
$T_P=4\Gamma^2|\boldsymbol u_D^TM^{-1}\boldsymbol u_S|^2$
and
\begin{equation}
\begin{split}
\left\langle\frac{G}{G_0}\right\rangle_P
\simeq{}&\frac{4\Gamma^2}{(2\pi)^{3/2}}
\int_{\mathbb R^3}d^3z\,e^{-|\boldsymbol z|^2/2}
\\
&\times
\left|\boldsymbol u_D^TM^{-1}\boldsymbol u_S\right|^2.
\end{split}
\label{eq:EM_pair_integral}
\end{equation}
This expression preserves $0\leq T_P\leq1$ and recovers
Eq.~\eqref{eq:EM_growth} at weak disorder. Projected
disorder scales as $\sigma/\sqrt N$, whereas contact
broadening scales as $\gamma/N$. Their characteristic
ratio,
\begin{equation}
\eta=\frac{\sigma\sqrt N}{\gamma},
\end{equation}
therefore increases with size, explaining the departure
from initial linear growth. The full inverse includes
repeated scattering within the pair; coupling through
other ring states is neglected.

\begin{figure}
{\centering \resizebox*{7.5cm}{3.6cm}{\includegraphics{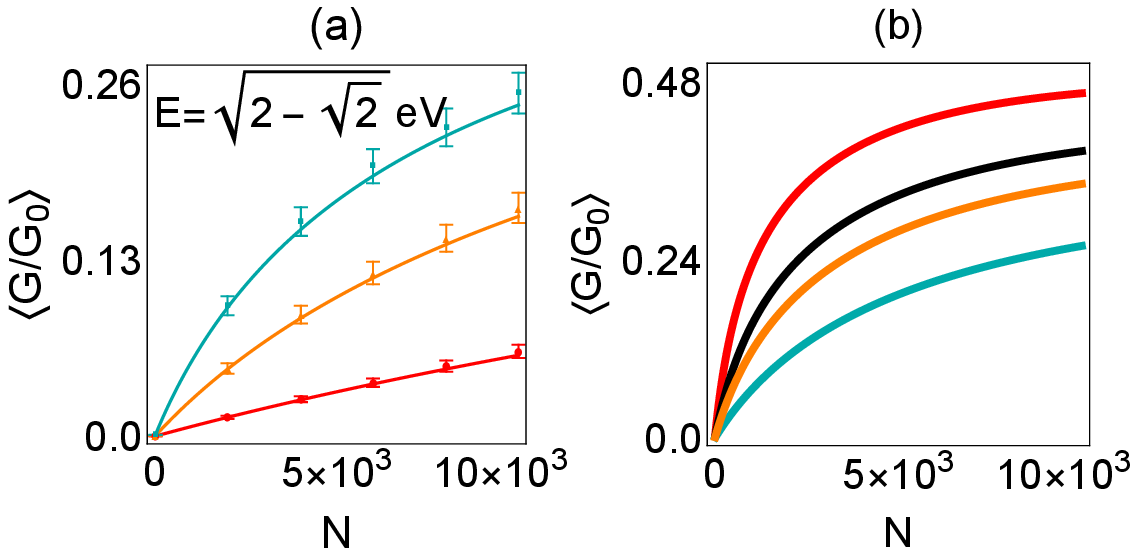}}\par}
\caption{(Color online) Disorder-averaged conductance
versus ring size for $N_D=N/2+2$, $v=1\,\mathrm{eV}$,
$t_0=4\,\mathrm{eV}$, and zero clean on-site energies.
Solid curves are parameter-free predictions of the
degenerate-pair expression, Eq.~\eqref{eq:EM_pair_integral}.
(a) $E_F=2v\cos(3\pi/8)=\sqrt{2-\sqrt2}\,\mathrm{eV}$
and $\tau_c=0.2\,\mathrm{eV}$, with
$\Delta W=0.1$, $0.2$, and $0.3\,\mathrm{meV}$
shown in red, orange, and teal, respectively.
Circles, triangles, and squares show the corresponding
numerical means at selected ring sizes. Error bars
indicate approximate $95\%$ confidence intervals for
the means, estimated from 1000 independent disorder
realizations.
(b) Analytical predictions at $\Delta W=0.1\,\mathrm{meV}$
and $\tau_c=0.1\,\mathrm{eV}$ for
$E_F=2v\cos(\pi/4)$ (red),
$2v\cos(7\pi/10)$ (black),
$2v\cos(\pi/3)$ (orange), and
$2v\cos(4\pi/7)$ (teal).
Ring sizes follow the corresponding commensurate
sequences with $N\leq10000$.}
\label{f6}
\end{figure}

\paragraph{Comparison and predictions.}
Figure~\ref{f6}(a) compares the parameter-free predictions
of Eq.~\eqref{eq:EM_pair_integral} with numerical disorder
averages at $E_F=\sqrt{2-\sqrt2}\,\mathrm{eV}$ and
$\tau_c=0.2\,\mathrm{eV}$. Analytical curves use
$N=16m$, $m=1,\ldots,625$, while numerical symbols
show selected sizes from the sequence $N=16+32r$.
Error bars indicate approximate $95\%$ confidence
intervals for the means, with half-width
$1.96s_g/\sqrt{1000}$, where $s_g$ is the sample
standard deviation of $g=G/G_0$ over 1000 independent
disorder realizations. The analytical treatment captures
the growth and increasing curvature with disorder
in Fig.~\ref{f4}(a).

At $N=10000$, the predictions are $0.05979$, $0.16230$,
and $0.24448$ for $\Delta W=0.1$, $0.2$, and
$0.3\,\mathrm{meV}$, respectively. The corresponding
numerical means are $0.06264\pm0.00250$,
$0.16864\pm0.00568$, and $0.25327\pm0.00768$,
where the quoted uncertainties denote one standard error.
The relative differences of $3.5$--$4.6\%$ are compatible
with sampling uncertainty: all three predictions lie
within the numerical $95\%$ confidence intervals.
Here $\eta=0.289$, $0.577$, and $0.866$, respectively,
motivating retention of the full denominator.

Figure~\ref{f6}(b) predicts growth at four additional
energies for $\Delta W=0.1\,\mathrm{meV}$ and
$\tau_c=0.1\,\mathrm{eV}$.
The wave numbers $k=\pi/4$, $7\pi/10$, $\pi/3$, and
$4\pi/7$ correspond to the commensurate sequences
$N=8\ell$, $20\ell$, $6\ell$, and $14\ell$, respectively.
Their growth rates and curvature depend on both the
contact phase and electrode self-energy.

A separate comparison with Fig.~\ref{f4}(c), at
$E_F=-2\cos(\pi/7)\,\mathrm{eV}$ and
$\tau_c=0.25\,\mathrm{eV}$, gives analytical predictions
$0.36282$, $0.54258$, and $0.61415$ at $N=9982$
for $\Delta W=0.1$, $0.2$, and $0.3\,\mathrm{meV}$,
respectively. The corresponding numerical means are
$0.36123\pm0.00913$, $0.53597\pm0.01001$, and
$0.60703\pm0.00976$, where uncertainties denote one
standard error over 1000 independent disorder realizations.
The relative differences are below $1.3\%$, and all three
predictions lie within the numerical $95\%$ confidence
intervals. Together, these comparisons support a common
mechanism: disorder lifts the clean transmission zero,
while repeated scattering modifies the initial growth.

\paragraph{Band-center exception and validity.}
At $E=0$, for $N=4\ell$ and odd $d$, the clean
contact columns occupy opposite sublattices, giving
$\mathcal G^{(1)}_{d0}=0$. The full-ring second-order
amplitude instead yields
\begin{equation}
\left\langle\frac{G(0)}{G_0}\right\rangle
=\frac{(\Delta W)^4[d^2+(N-d)^2-2]}
{288\gamma^2v^2}
+O((\Delta W)^6).
\label{eq:EM_center}
\end{equation}
For $N_D=N/2+2$,
\begin{equation}
\left\langle\frac{G(0)}{G_0}\right\rangle
=\frac{(\Delta W)^4N^2}{576\gamma^2v^2}
+O((\Delta W)^6).
\end{equation}
For $N_D=N/2+4$, $N^2$ is replaced by $N^2+32$.
The leading band-center growth is therefore quadratic.
Virtual transitions outside the zero-energy pair are
essential, precluding direct application of the
nonzero-energy pair approximation.

The perturbative expressions require small higher-order
corrections. The projected integral extends their range
by retaining repeated scattering within the selected
pair, provided contributions from other levels remain
weak. It describes the observed growth and curvature,
but does not determine the asymptotic conductance of
the full disordered ring.


\begin{thebibliography}{99}

\bibitem{r01}
P. W. Anderson, Phys. Rev. \textbf{109}, 1492 (1958).

\bibitem{r02}
E. Abrahams, P. W. Anderson, D. C. Licciardello,
and T. V. Ramakrishnan, Phys. Rev. Lett. \textbf{42}, 673 (1979).

\bibitem{r03}
P. A. Lee and T. V. Ramakrishnan,
Rev. Mod. Phys. \textbf{57}, 287 (1985).

\bibitem{r04}
B. Kramer and A. MacKinnon, Rep. Prog. Phys. \textbf{56}, 1469 (1993).

\bibitem{r05}
F. Evers and A. D. Mirlin, Rev. Mod. Phys. \textbf{80}, 1355 (2008).

\bibitem{r06}
D. J. Thouless, Phys. Rep. \textbf{13}, 93 (1974).

\bibitem{r07}
P. W. Anderson, D. J. Thouless, E. Abrahams, and D. S. Fisher,
Phys. Rev. B \textbf{22}, 3519 (1980).

\bibitem{r08}
M. Kappus and F. Wegner, Z. Phys. B \textbf{45}, 15 (1981).

\bibitem{r09}
H. Schomerus and M. Titov, Phys. Rev. B \textbf{67}, 100201(R) (2003).

\bibitem{r10}
L. I. Deych, A. A. Lisyansky, and B. L. Altshuler,
Phys. Rev. Lett. \textbf{84}, 2678 (2000).

\bibitem{r11}
L. I. Deych, M. V. Erementchouk, A. A. Lisyansky,
and B. L. Altshuler,
Phys. Rev. Lett. \textbf{91}, 096601 (2003).

\bibitem{r12}
C. W. J. Beenakker, Rev. Mod. Phys. \textbf{69}, 731 (1997).

\bibitem{r13}
A. A. Chabanov, M. Stoytchev, and A. Z. Genack,
Nature \textbf{404}, 850 (2000).

\bibitem{r14}
T. Schwartz, G. Bartal, S. Fishman, and M. Segev,
Nature \textbf{446}, 52 (2007).

\bibitem{r15}
Y. Lahini \textit{et al.},
Phys. Rev. Lett. \textbf{100}, 013906 (2008).

\bibitem{r16}
J. Billy \textit{et al.},
Nature \textbf{453}, 891 (2008).

\bibitem{r17}
G. Roati \textit{et al.},
Nature \textbf{453}, 895 (2008).

\bibitem{r18}
R. Landauer,
IBM J. Res. Dev. \textbf{1}, 223 (1957).

\bibitem{r19}
R. Landauer,
Philos. Mag. \textbf{21}, 863 (1970).

\bibitem{r20}
D. S. Fisher and P. A. Lee,
Phys. Rev. B \textbf{23}, 6851 (1981).

\bibitem{r21}
E. N. Economou and C. M. Soukoulis,
Phys. Rev. Lett. \textbf{46}, 618 (1981).

\bibitem{r22}
M. B\"uttiker,
Phys. Rev. Lett. \textbf{57}, 1761 (1986).

\bibitem{r23}
Y. Imry and R. Landauer,
Rev. Mod. Phys. \textbf{71}, S306 (1999).

\bibitem{r24}
P. A. Lee and A. D. Stone,
Phys. Rev. Lett. \textbf{55}, 1622 (1985).

\bibitem{n01}
C. J. Lambert,
Chem. Soc. Rev. \textbf{44}, 875 (2015).

\bibitem{n02}
F. Evers, R. Koryt\'ar, S. Tewari, and J. M. van Ruitenbeek,
Rev. Mod. Phys. \textbf{92}, 035001 (2020).

\bibitem{r26}
Y. Aharonov and D. Bohm,
Phys. Rev. \textbf{115}, 485 (1959).

\bibitem{r28}
M. B\"uttiker, Y. Imry, and M. Ya. Azbel,
Phys. Rev. A \textbf{30}, 1982 (1984).

\bibitem{r29}
Y. Gefen, Y. Imry, and M. Ya. Azbel,
Phys. Rev. Lett. \textbf{52}, 129 (1984).

\bibitem{r30}
R. A. Webb, S. Washburn, C. P. Umbach, and R. B. Laibowitz,
Phys. Rev. Lett. \textbf{54}, 2696 (1985).

\bibitem{r32}
A. Yacoby, M. Heiblum, D. Mahalu, and H. Shtrikman,
Phys. Rev. Lett. \textbf{74}, 4047 (1995).

\bibitem{r33}
R. Schuster \textit{et al.},
Nature \textbf{385}, 417 (1997).

\bibitem{n19}
Z. Geng \textit{et al.},
Phys. Rev. B \textbf{105}, L241112 (2022).

\bibitem{r34}
U. Fano,
Phys. Rev. \textbf{124}, 1866 (1961).

\bibitem{r35}
A. E. Miroshnichenko, S. Flach, and Y. S. Kivshar,
Rev. Mod. Phys. \textbf{82}, 2257 (2010).

\bibitem{r37}
K. Kobayashi, H. Aikawa, S. Katsumoto, and Y. Iye,
Phys. Rev. Lett. \textbf{88}, 256806 (2002).

\bibitem{r38}
K. Kobayashi, H. Aikawa, S. Katsumoto, and Y. Iye,
Phys. Rev. B \textbf{68}, 235304 (2003).

\bibitem{r41}
A. Aharony, O. Entin-Wohlman, B. I. Halperin, and Y. Imry,
Phys. Rev. B \textbf{66}, 115311 (2002).

\bibitem{r44}
H. Friedrich and D. Wintgen,
Phys. Rev. A \textbf{32}, 3231 (1985).

\bibitem{r46}
C. W. Hsu, B. Zhen, A. D. Stone, J. D. Joannopoulos,
and M. Solja\v{c}i\'c,
Nat. Rev. Mater. \textbf{1}, 16048 (2016).

\bibitem{r47}
I. Rotter,
J. Phys. A: Math. Theor. \textbf{42}, 153001 (2009).

\bibitem{n03}
J. Liu, X. Huang, F. Wang, and W. Hong,
Acc. Chem. Res. \textbf{52}, 151 (2019).

\bibitem{n04}
Y. Li \textit{et al.},
Nat. Mater. \textbf{18}, 357 (2019).

\bibitem{n05}
J. Bai \textit{et al.},
Nat. Mater. \textbf{18}, 364 (2019).

%\bibitem{n06}
%Z. Chen \textit{et al.},
%Nat. Nanotechnol. \textbf{19}, 986 (2024).

%\bibitem{n07}
%X. Xu \textit{et al.},
%Phys. Rev. B \textbf{112}, 155431 (2025).

\bibitem{n08}
S. Flach, D. Leykam, J. D. Bodyfelt, P. Matthies,
and A. S. Desyatnikov,
Europhys. Lett. \textbf{105}, 30001 (2014).

\bibitem{n09}
J. D. Bodyfelt, D. Leykam, C. Danieli, X. Yu, and S. Flach,
Phys. Rev. Lett. \textbf{113}, 236403 (2014).

\bibitem{n10}
C. Danieli, J. D. Bodyfelt, and S. Flach,
Phys. Rev. B \textbf{91}, 235134 (2015).

\bibitem{n11}
D. Leykam, J. D. Bodyfelt, A. S. Desyatnikov, and S. Flach,
Eur. Phys. J. B \textbf{90}, 1 (2017).

\bibitem{n12}
D. Leykam, A. Andreanov, and S. Flach,
Adv. Phys.: X \textbf{3}, 1473052 (2018).

\bibitem{n13}
S. Mukherjee, M. Di Liberto, P. \"Ohberg, R. R. Thomson,
and N. Goldman,
Phys. Rev. Lett. \textbf{121}, 075502 (2018).

\bibitem{n14}
S. Longhi,
Opt. Lett. \textbf{46}, 2872 (2021).

\bibitem{n15}
H. Li, Z. Dong, S. Longhi, Q. Liang, D. Xie, and B. Yan,
Phys. Rev. Lett. \textbf{129}, 220403 (2022).

\bibitem{r54}
H. Wang, W. Zhang, H. Sun, and X. Zhang,
Phys. Rev. B \textbf{106}, 104203 (2022).

\bibitem{n16}
W. Zhang, H. Wang, H. Sun, and X. Zhang,
Phys. Rev. Lett. \textbf{130}, 206401 (2023).

\bibitem{n18}
G. M. A. Almeida, R. F. Dutra, A. M. C. Souza, M. L. Lyra,
and F. A. B. F. de Moura,
Phys. Rev. A \textbf{108}, 022407 (2023).

\bibitem{n20}
M. Walschaers, R. Mulet, T. Wellens, and A. Buchleitner,
Phys. Rev. E \textbf{91}, 042137 (2015).

\bibitem{r55}
N. C. Ch\'avez \textit{et al.},
Phys. Rev. Lett. \textbf{126}, 153201 (2021).

\bibitem{r56}
T. F. Allard and G. Weick,
Phys. Rev. B \textbf{106}, 245424 (2022).

\bibitem{r57}
E. Zanardini, G. L. Celardo, N. C. Ch\'avez, and F. Borgonovi,
Phys. Rev. B \textbf{113}, 014202 (2026).

\bibitem{n17}
P. Alc\'azar Guerrero \textit{et al.},
Phys. Rev. Lett. \textbf{134}, 126301 (2025).

\bibitem{r48}
P. M. Shmakov, A. P. Dmitriev, and V. Yu. Kachorovskii,
Phys. Rev. B \textbf{87}, 235417 (2013).

\bibitem{r58}
M. B\"uttiker,
Phys. Rev. B \textbf{33}, 3020 (1986).

\bibitem{r59}
J. L. D\'{A}mato and H. M. Pastawski,
Phys. Rev. B \textbf{41}, 7411 (1990).

\bibitem{meV1} D. M. Zumb\"uhl, C. M. Marcus, M. P. Hanson, and A. C. Gossard,
Phys. Rev. Lett. \textbf{96}, 206802 (2006).

\bibitem{meV2} D. Goldhaber-Gordon, H. Shtrikman, D. Mahalu,
{\it et al.} Nature \textbf{391}, 156 (1998).

\bibitem{meV3} S. M. Cronenwett, T. H. Oosterkamp, and
L. P. Kouwenhoven, Science \textbf{281}, 540 (1998).

\end{thebibliography}
\end{document}